Alexander Barabanov, Maxim Grishin, Alexey Markov

# The Formal Metabasis For Conformity Assessment of Information Security Software and Hardware

**Abstract:** An approach to the development of security test procedures for information security controls is presented. The recommendations for optimizing the test procedure are obtained.

**Introduction**

The mandatory evaluation of the information security software and hardware (ISSH) used for informatization objects is performed as ISSH certification testing. The conformity evaluation is also performed at the stages of acceptance testing, informatization objects validation, and information security efficiency control. ISSH requirements are defined in the regulatory and procedural guidelines for regulators, but those, however, either lack the descriptions of testing methods or provide them in a qualitative description manner, and this hinders automation and optimization of ISSH conformity evaluation. This presentation discusses formalization of both general and particular methods of ISSH conformity evaluation so as to enable one to define time, cost, and exhaustiveness factors of ISSH testing.

**1. Formal Metabasis of Conformity Evaluation**

An ISSH set is a set of hardware and software data processing components acting either independently or as parts of other systems designed for either prevention or setting a considerable obstacle to unauthorized access to information [1]. ISSH includes a number of security subsystems, such as identification, authentication, access isolation, integrity control, logging, and other mechanisms designed to fight actual information security threats.

Let us discuss an ISSH test. Let us grant that $R = \{r_i\}$ is a subset of requirements for ISSH $\Sigma$ while $T = \{t_i\}$ is a set of processes used to test the conformity to the requirements.

The *test process designing method* will be described as the following transformation: $M: \Sigma \times R \to T$. Function $M$, on the basis of requirement $r_i \in R$ and information on the implementation of ISSH $\Sigma$ under test, generates test process $t_i \in T$ performed for testing the ISSH compliance with requirement $r_i \in R$. As a rule, function $M$ for ISSH $\Sigma$ under test is a bijection.

Any test process $t_i \in T$ is characterized by the following: the purpose, the workflow, the results to be registered, and the criteria of the positive decision.

*The purpose* contains the description of the intention to test ISSH conformity to the requirements. *The workflow* defines the set of instructions to be performed by the tester to initialize the ISSH under test and generate the input sequence to the ISSH. *The results of test processes* are registered with different software testing means, such as network traffic generation and interception tools, bulk memory search applications, and access isolation testing applications. *The criteria of the positive decision* must contain the test processes model results. The test is to compare the model and actual test results in order to enable a decision of the ISSH conformity or non-conformity.

Let us introduce operators of meeting the requirements $F_R$ and correctness of test results $F_C$ for the given $\Sigma, r_i, t_i$.

Operator of meeting requirement $r_i$ for ISSH $\Sigma$ $F_R: \Sigma \times R \to \{0,1\}$:

$$F_R(\Sigma, r_i) = \begin{cases} 1, & \text{if requirement } r_i \text{ holds for } \Sigma, \\ 0, & \text{if no.} \end{cases}$$

Operator of correctness of test process $t_i$ for ISSH $\Sigma$ $F_C: \Sigma \times T \to \{0,1\}$:

$$F_C(\Sigma, t_i) = \begin{cases} 1, & \text{if test } t_i \text{ successfully passed for } \Sigma, \\ 0, & \text{if no.} \end{cases}$$

Operator $F_C$ indicates successful or unsuccessful test results for ISSH $\Sigma$, that is, whether the actual test results are equivalent to the model results indicated in the process description.

Set of five objects $\langle \Sigma, R, M, F_R, F_C \rangle$ will be *the ISSH test method*, where $R$ is a set of requirements for ISSH $\Sigma$, $M$ is a test process designing method, and $F_R$ and $F_C$ are operators of meeting the requirements and correctness of test processes respectively, and for $\forall r_i \in R$ $F_R(\Sigma, r_i) = F_C(\Sigma, t_i)$ holds.

The method presupposes three stages: planning, testing, and result analysis. At the *planning* stage, the specifications and ISSH features are analyzed. Before testing, the testers must confirm that the ISSH specifications contain the creator's claim of ISSH conformity to requirements $R$, that is $F_R(\Sigma, r_i) = 1$ for $\forall r_i \in R$. On the basis of data obtained in specifications analysis and ISSH test runs, and the requirements, a set of test processes $T = \{t_i\}$ is generated, where $t_i = M(\Sigma, r_i)$. $T = \{t_i\}$ test processes are used for *testing* of the system. For each test process $t_i$, the results to be registered are obtained. At the *analysis* stage a set of ordered couples $(t_i, F_C(\Sigma, t_i))$ of actual and model results, is obtained. For ISSH $\Sigma$, the conformity to requirements $R = \{r_1, r_2, ..., r_i, ..., r_n\}$ is declared is:

$$\sum_{i=1}^{n} F_C(\Sigma, t_i) = \{r_1, r_2, ..., r_i, ..., r_n\},$$

that is, the test has proved the conformity of actual ISSH capabilities to the ones either claimed by the specifications or required by regulations.

**2. Methods of Testing of Information Security Tools Conformity to Information Security Requirements**

The basic document containing ISSH requirements is the regulatory document on computational technology products by the Russian State Technical Committee[1] [1]. By this document, 7 protection classes are established and requirements are formulated to discretionary and mandatory access controls, memory wiping, module isolation, document marking, protection of input and output to an alienated material carrier, user-device matching, identification and authentication, events logging, integrity control, etc. Let us discuss the formalized test procedures of the most resource-hungry requirements of the above $R_{IST} = \{r_1, r_2, r_3, r_4, r_5, r_6\}$ taken from this document.

**2.1. Particular Methods for Discretionary Access Isolation Testing**

The purpose is to test whether the ISSH functional opportunities meet the requirements for discretionary access isolation, and if yes, to what extent.

Let us introduce definitions to be used for description of the test process. Let us grant that $S = \{S_1, S_2, ..., S_i, ..., S_n\}$ is a set of test access subjects, $O = \{O_1, O_2, ..., O_j, ..., O_m\}$ is a set of test access objects, and $R = \{R_1, R_2, ..., R_k, ..., R_l\}$ is a set of possible access rights (e.g., view, saving, deletion, etc.). Let us define the access matrix as $M = (m_{ij})$, $m_{ij} \subseteq R$, where $m_{ij}$ is a set of access rights of the test subject $S_i$ to the test object $O_j$. The matrix row corresponds to the subject $S_i$, and the column to the object $O_j$. The element at the crossing of the row and column contains the set of access rights $m_{ij} \subseteq R$ of the corresponding subject to the corresponding object.

---
[1] www.fstec.ru

The operator of the possession by the subject of the right to access to the object in the matrix will be $F_M(S,O,R)$:

$$F_M(S_j, O_i, r_k) = \begin{cases} 1, & \text{if } r_k \in m_{ij} \\ 0, & \text{if } r_k \notin m_{ij} \end{cases}$$

The operator of the actual possession by the subject of the right to access to the object will be $F_{fa}(S,O,R)$:

$$F_{fa}(S_j, O_i, r_k) = \begin{cases} 1, & \text{if subject } S_j \text{ has right } r_k \text{ to object } O_i \\ 0, & \text{otherwise} \end{cases}$$

The sequence of the performed operations may be as follows:

1. Creation of test subjects $S = \{S_1, S_2, ..., S_n\}$ and objects of access $O = \{O_1, O_2, ..., O_m\}$. Testing will be performed for all possible subjects and objects of access, and the list of subjects and objects will be defined on the basis of the ISSH specifications analysis.

2. Adjustment of the access rights for subjects of the ISSH under test to the test objects under protection. This operation presupposes the adjustment of the access matrix $M = (m_{ij})$, $m_{ij} \subseteq R$. In the testing, all possible subject-to-object access rights and their combinations are checked.

3. Testing of ISSH configurations: the check of the actual possession of the right $r_k$ by the subject $S_j$ with relation to the object $O_i$; thus, all values of the operator $F_{fa}(S,O,R)$ are checked for any $i, j, k$. The check is performed with the test access attempts, such as viewing, saving, or deletion of objects by subjects.

4. Comparing the actual access rights with those defined by the access matrix.

The results of the test to be registered are:

1. The set of test access subjects $S = \{S_1, S_2, ..., S_n\}$, the set of test access objects $O = \{O_1, O_2, ..., O_m\}$, and the set of possible access rights $R = \{R_1, R_2, ..., R_p\}$.

2. The results of adjustment of access isolation rules, that is, access matrix $M = (m_{ij})$, $m_{ij} \subseteq R$.

3. Results of the check of the actual possession of the right $R_k$ by the subject $S_j$ with relation to object $O_i$, that is, values of the operator $F_{fa}(S,O,R)$ for all $i, j, k$.

Let us define the criterion of positive decision: as a result of comparison of actual and required access rights defined by the access matrix, they must coincide:

The check of the set access rights is performed through attempts of open and close access by the subjects to objects with registration of results of these attempts as successful or not.

The analysis of the obtained data is performed through comparison of the access attempts results with the expected results defined in the test access matrix.

Similar to discretionary access isolation testing, the *user-device matching testing* is performed, but in this case different input and output devices are considered access objects.

**2.2. A Particular Method of Mandatory Access Control Testing**

Let us introduce definitions to be used for description of the test process. Let us grant that $S=\{S_1, S_2, ..., S_n\}$ is a set of test access subjects, $O=\{O_1, O_2, ..., O_m\}$ is a set of test access objects, $M=\{m_1, m_2, ..., m_k\}$ is a set of classification labels (classification levels) of access subjects and objects (classification labels has an hierarchy: $m_1 < m_2 < ... < m_k$), $m_{Oi}$ is a classification label of the $i^{th}$ access object and $m_{Sj}$ is classification label of the $j^{th}$ access subject.

Let us introduce the operators $F_{READ}(S_i, O_j) = \{0, 1\}$ of checking the right to reading and $F_{WRITE}(S_i, O_j) = \{0, 1\}$ of checking the right to writing possessed by subject $S_i$ with relation to object $O_j$:

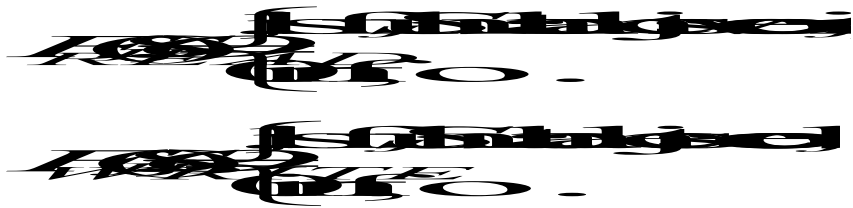

The sequence of the performed operations may be as follows:

1. Creating test access subjects $S=\{S_1, S_2, ..., S_n\}$ and objects $O=\{O_1, O_2, ..., O_m\}$.

2. Assigning classification labels $M=\{m_1, m_2, ..., m_k\}$ to the access subjects (by transformation $F_S: S \to M$ which permits to calculate the classification level of any access subject, i. e. $F_S(S_i) = m_{Si}$).

3. Assigning classification labels $M=\{m_1, m_2, ..., m_k\}$ to the access objects (by transformation $F_O: O \to M$ which permits to calculate the classification level of any access object $F_O(O_j) = m_{Oj}$).

4. Performing the following test object access attempts by the subjects:

– reading data of the access objects by the access subjects, i. e. calculation of $F_{READ}(S,O)=1$ for $\forall i,j$;

– writing data to the access objects by the access subjects, i. e. calculation of $F_{WRITE}(S,O)=1$ for $\forall i,j$.

5. Checking if the obtained results conform to the rules of the ticket-oriented access isolation.

The results of the test to be registered are:

1. The set of test access subjects $S=\{S_1, S_2, ..., S_n\}$, the set of test access objects $O=\{O_1, O_2, ..., O_m\}$, the set of classification labels (classification levels) of access subjects and objects $M=\{m_1, m_2, ..., m_k\}$.

2. The results of adjustment of access isolation rules, that is, transformations $F_S: S \to M$, $F_O: O \to M$ (classification labels of access subjects and objects).

3. Results of the check of the actual possession of the right to writing and reading by the subject $S_j$ with relation to the object $O_i$, that is, values of operators $F_{READ}(S,O)=1$ and $F_{WRITE}(S,O)=1$ for $\forall i,j$.

As a result of the check of the ticket-oriented access isolation rules, the following data are obtained:

– the access subject $S_j$ of the classification level $m_{Sj}$ may read data of the access object $O_i$ of the classification level $m_{Oi}$ if and only if the classification level of the access subject $S_j$ is either higher or equal to the classification level of the access object $O_i$, i. e. for $\forall i,j$, $F_{READ}(S,O)=1$ if and only if $m_{Sj} \geq m_{Oi}$;

– the access subject $S_j$ of the classification level $m_{Sj}$ may write data to the access object $O_i$ of the classification level $m_{Oi}$ if and only if the classification level of the access object $O_i$ is either higher or equal to the classification level of the access subject $S_j$, i. e. for $\forall i,j$, $F_{WRITE}(S,O)=1$ if and only if $m_{Oi} \geq m_{Sj}$.

In a similar way, the test of the *protection of input and output to an alienated material carrier* is performed.

## 2.3. A Particular Method of Memory Cleaning Testing

Let us introduce definitions to be used for description of the test process. Let us grant that $A=\{a_1, a_2, ..., a_n\}$ is a set of memory areas under test (short term memory, hard drive partitions, external carriers, etc.), and $S$ is a test sequence of symbols unique for each area A of the memory. For the description of the testing process, we will use the operator of presence of the test sequence in the memory $F_{check}(a_j, S)$:

$$F_{check}(a_j, S) = \begin{cases} 1, \text{if sequence } S \text{ is present in area } a_j, \\ 0, \text{if no.} \end{cases}$$

The sequence of operations is as follows:
1. Adjustment of memory cleaning applications.
2. Placing the test data into the memory (a unique sequence of text symbols $S$).
3. Locating the test data within the memory (address, disc partition, etc.).
4. Memory relocation (release) with the use of the standard bench means.
5. Control of the test data being present in or absent from the memory (a repeat search and addressing using addresses defined at the initial stage); defining the value of $F_{check}(a_j, S)$.
6. Results analysis.
7. Applying assessment criteria.

The tests results to be registered are as follows:
1. Results of memory cleaning applications adjustment.
2. Results of the check for the test data being present in the memory after relocation (release).

The criterion of positive decision is as follows: the sequence $S$ of symbols loaded into the memory must not be found again after relocation (release), i. e. $F_{check}(a_j, S) = 0$.

## 2.4. A Particular Method of Module Isolation Testing

The isolation of modules follows from the fact that each process run by a user has its individual addressing space isolated from any other process run by other users.

The sequence of operations is as follows:
1. Running applications or processes on behalf of different users.
2. Access attempts to the process memory run by a user on behalf of the user.
3. Access attempts to the process memory run by a user on behalf of the other users.
4. Access attempts to the PC real memory.

5. Results analysis.
6. Applying assessment criteria.

The test results to be registered are facts of access attempts to the processes' memory run by other users and access attempts to the PC real memory.

The criterion of the positive decision is as follows: within the test, no access to the processes' memory run by other users or to the PC real memory occurred.

## 2.5. A Particular Method of Access Subject Identification and Authentication Testing

Let us introduce definitions to be used for description of the test process. Granting that $A$ is the alphabet of passwords and IDs of ISSH users, the user will be designated as $id \in ID \subseteq A^*$, and the password as $pwd \in PWD \subseteq A^*$; the user's account $usr_i \in USR$ will be characterized by the sequence $usr_i = (id_i, pwd_i)$. Let us introduce the operator of correctness of authentication data $F_A: USR \to \{0,1\}$:

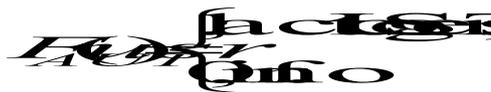

The following sequence of operations may be used for testing identification and authentication tools correctness:

1. Addressing the ISSH identification and authentication means and creating a set of access subject's accounts $USR = \{usr_i\}$.

2. Running queries for identification and authentication with the use of different combinations of authentication data: registered or unregistered IDs, or true or false passwords $try_j = (id_j, pwd_j)$.

3. Obtained data analysis.

The test results to be registered are:

1. ISSH configurations: the set of access subjects' accounts $USR$.

2. Data obtained in test queries for identification and authentication: the set $\{F_A(try_1), F_A(try_2), \ldots\}$.

The positive decision criteria are:

1. On entry of the registered ID and password, the user is granted an access to the protected data: $F_A(try) = 1, try \in USR$.

2. On entry of an unregistered ID and/or wrong password, the user is not granted an access to the protected data: $F_A(try) = 0, try \notin USR$.

The check of the reliability of the user identification and authentication processes' connection with all operations performed by the user is done within the tests of discretionary and ticket-based access control. After each check, the logs of events are analysed.

The check is considered a success if the audit log contains records about all access attempts and all operations performed by all users (security administrators included) that occurred within the tests described in the above paragraphs. In addition, each logging-in entry must specify the user ID employed in the access attempt and/or under which the user was logged on and performed different operations in the system.

**2.6. A Particular Method of Integrity Control Testing**

Granting that $FILES = \{file_1, file_2, ..., file_n\}$ is a set of ISSH files (configuration files or software module), let us introduce operators of the integrity violation $F_{MOD}$ and ISSH file integrity control $F_{INT}$.

The operator of integrity violation $F_{MOD}(FILES)$:

$$F_{MOD}(file) = \begin{cases} 1, & \text{file integrity violated in testing,} \\ 0, & \text{if no.} \end{cases}$$

The operator of ISSH file integrity control $F_{INT}(FILES)$:

$$F_{INT}(file) = \begin{cases} 1, & \text{file integrity,} \\ 0, & \text{if no.} \end{cases}$$

Let us designate the set of ISSH files modified within the test as $FILES^\Delta = \{file_1^\Delta, file_2^\Delta, ..., file_n^\Delta\}$, granting that the file $file_i$ is transformed into $file_i^\Delta$. When ISSH integrity control is tested, the following sequence of operations may be performed:

1. Configuration of integrity control applications (reaction to integrity violation, integrity control method, check-out period, test conditions, etc.), and identification of the ISSH file set $FILES\{file_i\}$.

2. Introducing changed into ISSH files (configuration changes, substitution or modification of the executable files, etc.), the result being the set of modified files $FILES^\Delta\{file_i^\Delta\}$.

3. ISSH file integrity check initialization (creation of conditions under which the ISSH performs integrity control).

4. Analysis of the ISSH reaction to its software or data integrity violation.

The test results to be registered are:

1. ISSH file set $FILES\{file_i\}$.

2. Modified ISSH file set $FILE_M = \{file_i\}$.

3. ISSH reaction to integrity violation: $F_{INT}(FID_i) \in \{F_{INT}\}$

The criterion of positive decision is: the ISSH reveals all facts of integrity violation:
$F_{INT}(FID_i) = F_{\nu}(fid_i)$ for $\forall i \in [1,n]$.

## 3. Recommendations for Testing Processes Optimization

Our experience of conducting different ISTs tests by our accredited test laboratory shows that the main issue is the growth of time and material expenditures on standard operations while the newly developed ISSH complexity grows along with the number of platforms and environments where these ISTs can operate. For example, for a discretionary access isolation testing, one will need to perform $|S| \cdot |O| \cdot |R|$ standard operations.

In general, it can be shown that the time $\tau_\Sigma$ needed for a test grows exponentially: $\tau_\Sigma \sim n \cdot v^w$, where $n$ is the number of requirements under test, $w$ is the number of test units (for example, a user account), and $v$ is the number of possible values that the factor under test may have.

The problem of ISSH testing optimization may be formulated as follows. Granting that $\tau : T \times \Sigma \to \mathbb{N}_0$ is the time needed for experts to test the ISSH $\Sigma$ with the use of the testing process $t_i$ and the transformation $C : R \times \Sigma \to \mathbb{N}_0$ reflects the expenditures of the ISSH $\Sigma$ testing against requirements $R$. The optimization (minimization of time within a certain expenditures limit) will be as follows:

$$\begin{cases} \sum_i \tau(t_i, \Sigma) \to \min, \\ \sum_i C(r_i, \Sigma) \leq C_M. \end{cases}$$

where $C_M$ is the limit set to expenditures.

As methods permitting to optimize testing, the following ways may be suggested.

1. *Combining different test types.* When an ISSH is tested, we recommend performing certain tests simultaneously. For example, the events logging subsystem testing can be combined with access isolation control and identification/authentication testing.

2. *Testing with combinatorial overlapping methods.* Testing all possible combinations of input effects to check the software behaviour in full, is very labour-consuming. The combinatorial overlapping testing is an approach that not only decreases the costs and increases the efficiency of tests, but also reduces the probability of error in the software. The idea is generally as follows: a failure of software is in the majority of cases is not caused by an isolated incorrect input parameter but by a combination of two or more input parameters (from two to six parameters, as follows from empirical data). The combinatorial overlapping may be practicable for both configuration and input parameters testing.

3. *Using software testing tools.* To reduce time expenditures, software should be used that permits to automate testing. Both ready-made software products presented in the software market of today and script-based home-made programmes may be used [2,3].

**Conclusion**

The formalization of the ISSH security subsystems testing methods and processes presented in this paper will make automation of ISSH and secured products conformity assessment easier.

The main problem that companies encounter when testing the products is the increase in time and cost expenditures caused by a large number of similar-type tests of the object's functions to be performed on all possible input data domains. The suggested methods will permit to reduce expenditures on conformity assessment and solve the problem of assessment time minimization within a certain expenditure limit.

The conceptual approach to the conformity assessment formalization may be recommended to anyone performing all types of tests of hardware, software, and secured systems [4-10].

**References**

1. Legal and Technological Bases of Information Security of Automated and Single-user Computational Systems / Kotenko I. V., Kotukhov M. M., Markov A. S., et al. St. Petersburg: VUS, 2000. 190 p.

2. Barabanov A. A. Information Security Testing Tools // Information Security. Inside Publishers. 2011. No. 1. p. 2-4. – in Russian

3. Markov A. S., Mironov S. V., Tsirlov V. L. Experience of Network Insecurity Scanners Testing // Informational Fighting Against Terrorist Threat, 2005. No. 5. p. 109-122.

4. Barabanov A. V., Markov A. S., Fadin A. A. Software Certifying without Source Texts // Open Source Systems. SUBD. 2011. No. 4. p. 38-41. URL: http://www.npo-echelon.com/doc/software_certification.pdf

5. Markov A. S., Mironov S. V., Tsirlov V. L. Revealing Insecurities in Source Code. // Open Source Systems. SUBD, 2005. No. 12. p. 64-69. URL: http://www.osp.ru/os/2005/12/380655/

6. Markov A. S., Maslov V. G., Tsirlov V. L., Oleksenko I. A. Software Testing Against Security Requirements // Engineering Physics Institute Proceedings. 2009. No. 2 (12) p. 2-6.

7. A Time-saving Approach in Testing for Software Development / D. Yu. Burlak, A. V. Sedakov, A. N. Sudarenko, P. S. Sokolov // Engineering Physics Institute Proceedings. 2010. No. 2 (16) p. 32-39.

8. Gourlay J. S. A Mathematical Framework for the Investigation of Testing // IEEE Transactions on Software Engineering. 1983, Vol. SE-9, №. 6. P. 686-709.

9. Kuhn R., Kacker R., Lei Y. Practical Combinatorial Testing // NIST Special Publication 800-142. Washington: U.S. Government Printing Office, 2010. 75 p.

10. Tian J. Software Quality Engineering: Testing, Quality Assurance, and Quantifiable Improvement. Wiley, 2005. 440 p.